\documentclass[aip, jcp, amsmath, amssymb, reprint]{revtex4-1}
\usepackage{graphicx}% Include figure files
\usepackage{dcolumn}% Align table columns on decimal point
\usepackage{bm}% bold math
\usepackage{color}

\begin{document}

%\preprint{AIP/123-QED}

\title{Glass formation in a mixture of hard disks and hard ellipses}

\author{Wen-Sheng Xu}
\email{wsxu0312@gmail.com}
\affiliation{James Franck Institute, The University of Chicago, Chicago, Illinois 60637, USA}

\author{Xiaozheng Duan}
\email{xzduan@ciac.ac.cn}
\author{Zhao-Yan Sun}
\email{zysun@ciac.ac.cn}
\author{Li-Jia An}
\email{ljan@ciac.ac.cn}
\affiliation{State Key Laboratory of Polymer Physics and Chemistry, Changchun Institute of Applied Chemistry, Chinese Academy of Sciences, Changchun 130022, People's Republic of China}

\date{\today}% It is always \today, today, but any date may be explicitly specified

\begin{abstract}
We present an event-driven molecular dynamics study of glass formation in two-dimensional binary mixtures composed of hard disks and hard ellipses, where both types of particles have the same area. We demonstrate that characteristic glass-formation behavior appears upon compression under appropriate conditions in such systems. In particular, while a rotational glass transition occurs only for the ellipses, both types of particles undergo a kinetic arrest in the translational degrees of freedom at a single density. The translational dynamics for the ellipses is found to be faster than that for the disks within the same system, indicating that shape anisotropy promotes the translational motion of particles. We further examine the influence of mixture's composition and aspect ratio on the glass formation. For the mixtures with an ellipse aspect ratio of $k=2$, both translational and rotational glass transition densities decrease with increasing the disk concentration at a similar rate and hence, the two glass transitions remain close to each other at all concentrations investigated. By elevating $k$, however, the rotational glass transition density diminishes at a faster rate than the translational one, leading to the formation of an orientational glass for the ellipses between the two transitions. Our simulations imply that mixtures of particles with different shapes emerge as a promising model for probing the role of particle shape in determining the properties of glass-forming liquids. Furthermore, our work illustrates the potential of using knowledge concerning the dependence of glass-formation properties on mixture's composition and particle shape to assist in the rational design of amorphous materials.
\end{abstract}


%\pacs{Valid PACS appear here}% PACS, the Physics and Astronomy Classification Scheme.
%\keywords{Suggested keywords}%Use showkeys class option if keyword display desired

\maketitle

\section{Introduction}

Despite the fact that glasses are ubiquitous in nature and in our daily life, the physical mechanism of the glass transition remains one of the most fascinating questions in condensed matter physics.~\cite{RMP_83_587, JCP_137_080901, JCP_138_12A301} Glass formation occurs in a variety of materials, ranging from van der Waals liquids and polymers to colloids and granular matter. Assemblies of spheres provide a prototypical model for exploring the essential aspects of the liquid-glass transition, and much of the physics has been derived from the studies of spherical particles, e.g., from the investigations for colloidal and granular glasses (see a recent review in Ref.~\citenum{RPP_75_066501}). Going beyond the spherical particle paradigm, recent great interest has been focusing on probing the glass formation in systems composed of anisotropic particles.~\cite{PRL_94_215701, JCP_123_204505, JCP_126_014505, JCP_130_244906, JCP_130_244907, PRE_86_061503, PRL_105_055702, JCP_134_014503, PRL_110_015901} From the fundamental point of view, shape anisotropy is expected to induce novel dynamical phenomena and a better understanding of the glass transition can benefit from exploring the translation-orientation interplay in such systems. Investigations for glass formation in anisotropic particle systems are also of practical importance since recent theories~\cite{JCP_126_014505, JCP_130_244906, JCP_130_244907, PRE_86_061503} and experiments~\cite{PRL_105_055702, JCP_134_014503} have demonstrated that particle shape alone offers a promising route for realizing new kinetically arrested states with distinctive mechanical properties. 

A system of hard ellipsoidal particles is among the simplest models of anisotropic particles. The phase behavior has been explored extensively by various theoretical approaches and simulations in both two-dimensional hard ellipses~\cite{MP_63_623, PRA_39_6498, PRA_44_5306, JCP_106_2355, JCP_56_4729, PRA_42_2126, JCP_139_024501, JCP_140_124901, JCP_140_204502} and three-dimensional hard ellipsoids.~\cite{PRL_52_287, MP_55_1171, PRL_92_255506, PRE_75_020402, JCP_131_164513, JCP_134_201103, JCP_136_134505, JCP_138_064501} Recently, colloidal ellipsoids have been employed to experimentally explore a number of novel physical phenomena induced by particle shape.~\cite{Science_314_626, PRE_80_011403, JCP_133_124509, Nature_476_308, PRE_83_011403, PRL_108_228303, PRL_110_035501, PRL_107_065702, NC_5_3829, PRL_110_188301, PNAS_111_15362} Although the dynamics in monodisperse hard ellipsoidal particles at low and moderate densities has been investigated in recent simulations,~\cite{JCP_139_024501, PRE_85_061707, PRL_98_265702} much less information is available concerning the glassy dynamics of hard ellipses~\cite{PRL_107_065702, NC_5_3829, PRL_110_188301, PNAS_111_15362, SM_11_627} and hard ellipsoids.~\cite{EPL_84_16003} The ideal glass transitions in hard ellipsoids of evolution (i.e., hard uniaxial ellipsoids) have theoretically been treated in the framework of molecular mode-coupling theory (MCT).~\cite{PRE_62_5173} Molecular MCT predicts that an orientational glass, in which kinetic arrest appears in the rotational degrees of freedom but the system remains ergodic in the translational degrees of freedom, can be formed in hard ellipsoids with large elongations. This implies that a system of hard ellipsoidal particles with sufficiently large elongations exhibits two glass transitions: The first corresponds to the kinetic arrest in the rotational degrees of freedom and the second to the translational freezing of particles. It is only very recently that the formation of an orientational glass and the two glass transitions have been observed in experiments for monolayers of colloidal ellipsoids with large aspect ratios.~\cite{PRL_107_065702, NC_5_3829} For hard ellipsoids with small aspect ratio, however, a single glass transition is predicted by molecular MCT,~\cite{PRE_62_5173} a prediction that is confirmed by recent experiments~\cite{NC_5_3829, PRL_110_188301}and simulations.~\cite{EPL_84_16003, SM_11_627} Moreover, molecular MCT suggests that the translational glass transition point in hard ellipsoids of evolution displays a nonmonotonic variation with increasing aspect ratio.~\cite{PRE_62_5173} Such nonmonotonic behavior is supported by the diffusive properties of monodisperse hard ellipsoids, as shown in recent molecular dynamics (MD) simulations.~\cite{PRL_98_265702} Similar nonmonotonic behavior for the jamming transition point has been reported in numerical studies of hard ellipsoidal particles~\cite{Science_303_990, PRE_75_051304} and soft ellipses.~\cite{SM_6_2960, PRE_86_041303} In spite of these recent advances, understanding of glassy dynamics is far from complete even in two-dimensional hard ellipses.

The present paper focuses on the glass formation in binary mixtures composed of hard disks and hard ellipses, where both types of particles have the same size, as measured by their areas. Such mixtures offer the potential to better understand the influence of composition and particle shape on the glass formation, since crystallization is readily avoided by the anisotropic shape of the ellipses. The study is also relevant to colloidal experiments since real synthesized colloidal particles have polydispersity not only in size but also in shape. Using event-driven molecular dynamics (EDMD) simulations, we demonstrate that characteristic glass-formation behavior appears upon compression in the translational degrees of freedom for both types of particles and in the rotational degrees of freedom for the ellipses at appropriate compositions and aspect ratios. We analyze the glassy dynamics for both species within the same system and examine the influence of composition and aspect ratio on the glass formation. Both types of particles are found to undergo a kinetic arrest in the translational degrees of freedom at a single density, as determined from the MCT power-law fits, although the translational dynamics for the ellipses is faster than that for the disks when comparing their dynamics within the same system. The rotational glass transition for the ellipses sets in at a density close to the transitional one when the aspect ratio $k$ of the ellipses is equal to two, a result that is in agreement with recent experiments~\cite{NC_5_3829, PRL_110_188301} for monolayers of colloidal ellipsoids and simulations for hard ellipses.~\cite{NC_5_3829, SM_11_627} We further show that altering mixture's composition significantly affects both glass transition densities when the ellipse aspect ratio is fixed. Nevertheless, the density range between the two glass transitions appears to be independent of composition, because both glass transition densities decrease at a similar rate as the disk concentration increases. By elevating the aspect ratio $k$ of the ellipses, however, the rotational glass transition density diminishes at a faster rate than the translational one, leading to the formation of an orientational glass for the ellipses between the two transitions, in line with recent experiments for monolayers of colloidal ellipsoids~\cite{PRL_107_065702, NC_5_3829} and MCT predictions.~\cite{PRE_62_5173} Therefore, our simulations imply that mixing particles with different shapes provides a promising way for investigating the role of particle shape in determining the properties of glass-forming liquids. Furthermore, our work illustrates the potential of using knowledge concerning the dependence of glass-formation properties on mixture's composition and particle shape to assist in the rational design of amorphous materials.

The remainder of the paper is organized as follows. Section II provides details of our model system and simulations. Section III begins with a demonstration of appearance of characteristic glassy dynamics upon increasing density in a mixture of hard disks and hard ellipses, and then examines in detail the influence of mixture's composition and aspect ratio on the glass formation. Our findings are finally summarized in Section IV.

\section{Model and simulation details}

\begin{figure}[tb]
	\centering
	\includegraphics[angle=0,width=0.4\textwidth]{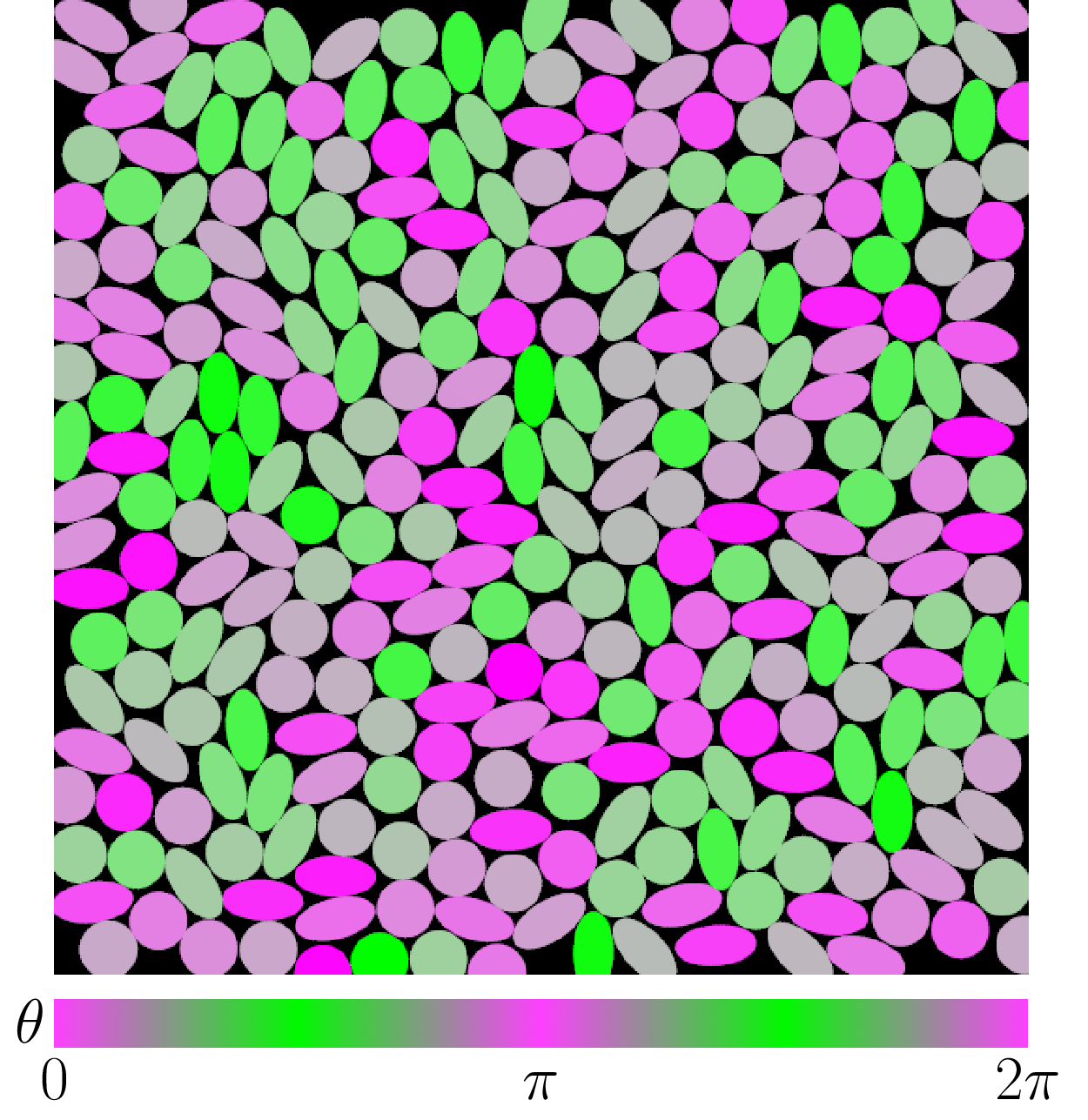}
	\caption{Typical snapshot for a mixture of hard disks and hard ellipses with the disk concentration $x=0.5$ and the ellipse aspect ratio $k=2$ at a high area fraction of $\phi=0.82$. Particles are colored according to their orientation, and $\theta$ denotes the angle between the particle semi-major axis and the $x$-axis.}
\end{figure}

The model system considered in the present paper is a two-dimensional binary mixture, composed of hard disks (denoted by $D$) and hard ellipses (denoted by $E$). All disks have an identical diameter $\sigma$. The aspect ratio of the ellipses is defined as $k=a/b$, where $a$ and $b$ denote the semi-major and semi-minor axes, respectively. The particle areas for the two components are chosen to be equal (i.e., $\sigma^2=4ab$) in the present work. Although other choices are also possible to avoid the crystallization, the present one allows us to better understand the influence of particle shape on the glassy dynamics, in particular within the same system, since the dynamics for different types of particles is compared at the same particle area. Other parameters that specify the system include the number concentration of the disks [$x=N_D/(N_D+N_E)$, where $N_{\alpha}$ (${\alpha}=D$ or $E$) is the number of ${\alpha}$ particles], and the area fraction (or density) of the system [$\phi=\pi(N_{D}\sigma^2/4+N_{E}ab)/L^2$ with $L$ the box dimension]. Figure 1 provides an illustrative snapshot for the model system with $x=0.5$ and $k=2$ at a high area fraction of $\phi=0.82$.

Although the disks possess an isotropically spherical shape, it is more convenient to assign angular velocities to all particles, since the system can then be regarded as a binary mixture of hard ellipses with $k=1$ for one component and $k>1$ for another. Hence, both types of particles move in both the translational and the rotational degrees of freedom in our simulations. EDMD simulations are performed in a square box under periodic boundary conditions. The reader is referred to Refs.~\citenum{JCP_139_024501, CDMD1, CDMD2} for details about implementing EDMD simulations for hard ellipses. All particles have the same mass $m$ and the same moment of inertia $I$. We set $m=I=1$ for convenience since the present paper only focuses on the general trends of static and dynamic quantities in variation with $\phi$, $x$ or $k$. The temperature $T$ is irrelevant for a hard-particle system and remains constant due to the conservation law of the total kinetic energy. We set the temperature as $k_BT=1$ for simplicity. Length and time are reported in units of $\sigma$ and $\sqrt{m\sigma^{2}/k_{B}T}$. The results reported in the present paper use the system with a total particle number of $N=N_D+N_E=300$. We also performed simulations for systems with $N=100$ and $N=500$ for selected state points. Finite size effects are found to be small for the properties considered using the present system size, and the conclusions given in the present paper are unaffected by the system size. A starting configuration with desired $\phi$, $x$ and $k$ is created using an extended Lubachevsky-Stillinger compression algorithm.~\cite{LS1, LS2, CDMD1, CDMD2} This starting configuration is then equilibrated for at least several relaxation times (defined in Subsection III A) before collecting data. For each state point, two or four independent samples are used to improve the statistics.

\section{Results and discussion}

This section first illustrates the appearance of typical glassy phenomena upon increasing density in mixtures of hard disks and hard ellipses. Specially, we analyze the structure and the dynamics for a system where both the disk concentration and the ellipse aspect ratio are fixed. The influence of mixture's composition and ellipse aspect ratio on the glass formation is then examined in detail.

\subsection{Glassy behavior in a mixture of hard disks and hard ellipses}

\begin{figure}[tb]
	\centering
	\includegraphics[angle=0,width=0.45\textwidth]{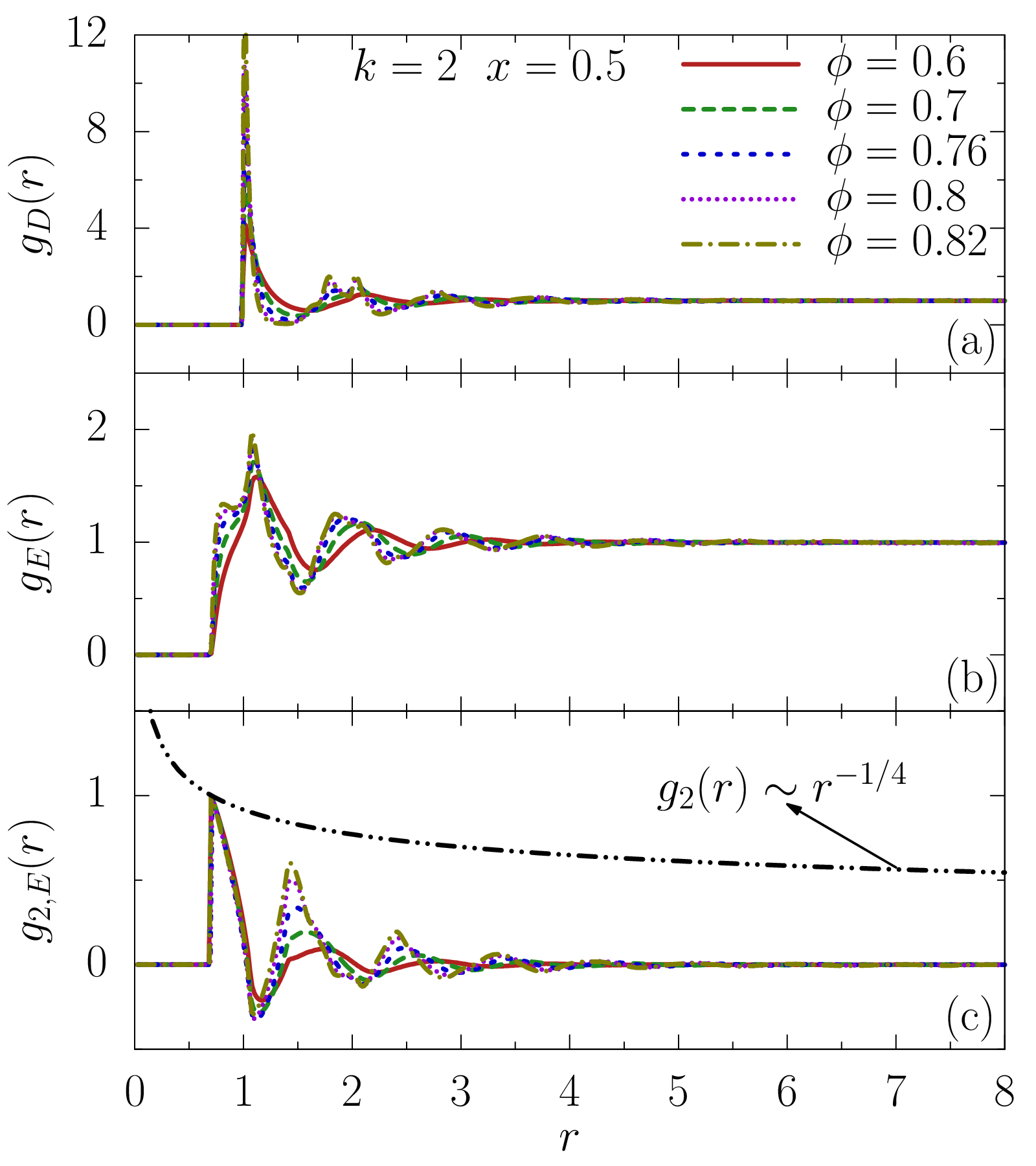}
	\caption{Pair correlation function $g_{\alpha}(r)$ for (a) the disks and (b) the ellipses, and (c) angular correlation function $g_{2,E}(r)$ for the ellipses for mixtures of hard disks and hard ellipses with $x=0.5$ and $k=2$ at various $\phi$ [indicated in (a)]. The power law $g_{2}(r)\sim r^{-1/4}$ is included in (c) to indicate the absence of any long-range orientational order in the system (see text).}
\end{figure}

\begin{figure}[tb]
	\centering
	\includegraphics[angle=0,width=0.45\textwidth]{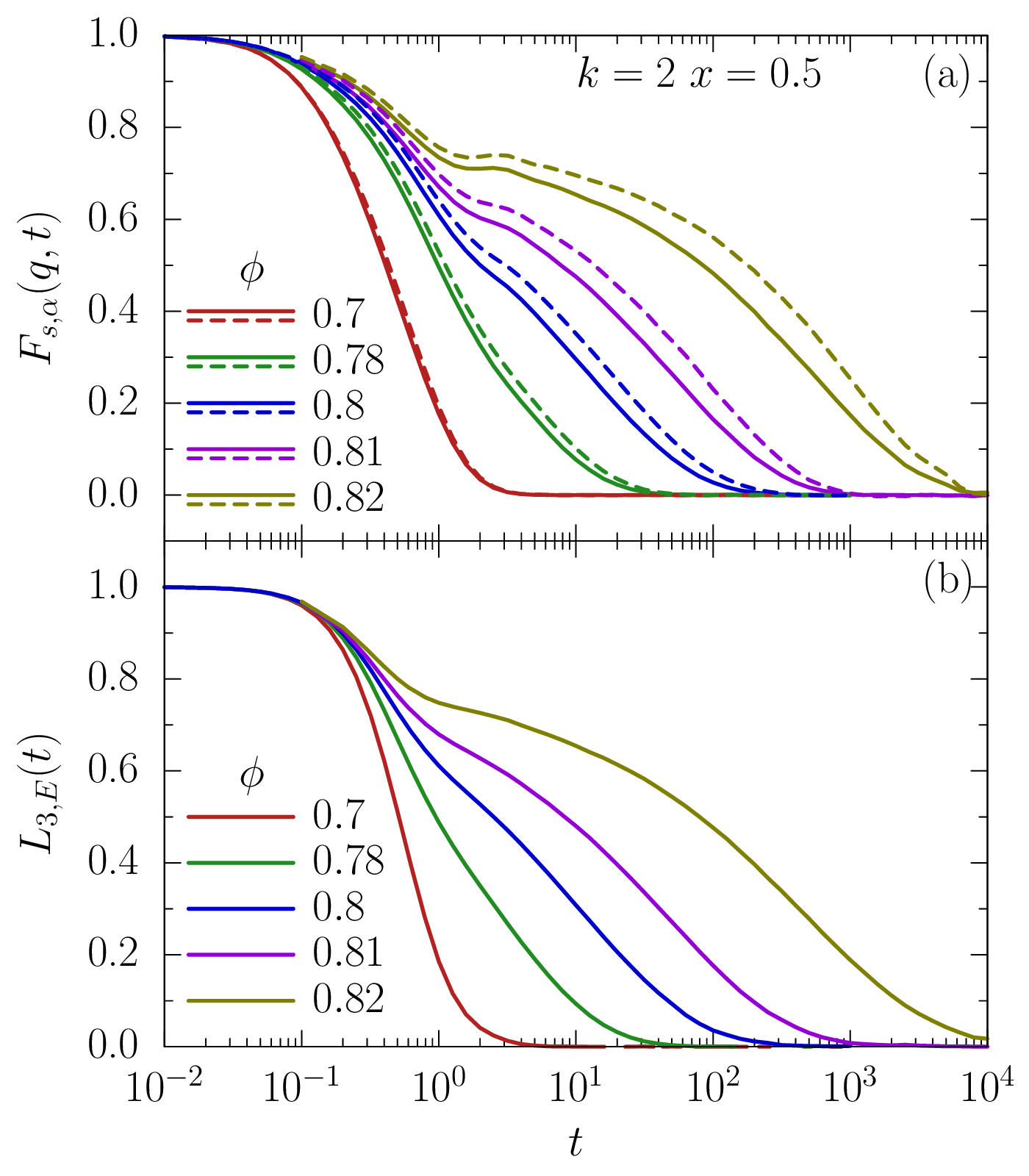}
	\caption{(a) Self-intermediate scattering function $F_{s,\alpha}(q,t)$ at $q=7$ for the disks (dashed lines) and the ellipses (solid lines) and (b) $3$rd order orientational correlation function $L_{3,E}(t)$ for the ellipses for mixtures of hard disks and hard ellipses with $x=0.5$ and $k=2$ at various $\phi$.}
\end{figure}

\begin{figure}[tb]
	\centering
	\includegraphics[angle=0,width=0.45\textwidth]{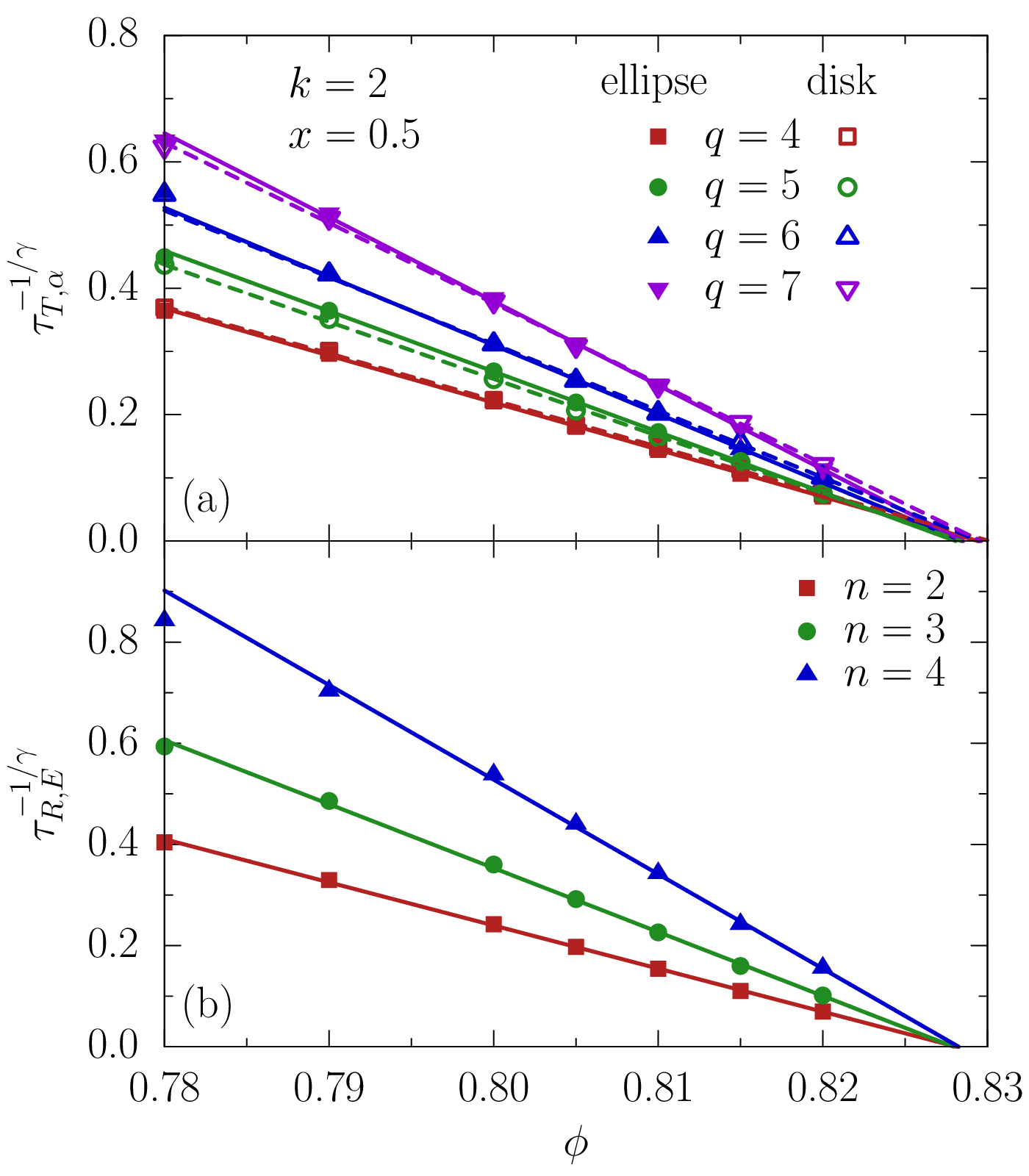}
	\caption{(a) $\tau_{T, \alpha}^{-1/\gamma}$ and (b) $\tau_{R, E}^{-1/\gamma}$ as a function of $\phi$ for mixtures of hard disks and hard ellipses with $x=0.5$ and $k=2$. Solid and dashed lines are MCT fits for the ellipses and the disks, respectively.  The fitted results are as follows. In (a): $(\gamma, \phi_{c, T_E})=(2.95, 0.8294)$ and $(\gamma, \phi_{c, T_D})=(3.25, 0.8301)$  for $q=4$, $(\gamma, \phi_{c, T_E})=(2.94, 0.8281)$ and $(\gamma, \phi_{c, T_D})=(3.13, 0.8284)$  for $q=5$, $(\gamma, \phi_{c, T_E})=(3.00, 0.8285)$ and $(\gamma, \phi_{c, T_D})=(3.33, 0.8295)$  for $q=6$, $(\gamma, \phi_{c, T_E})=(3.10, 0.8285)$ and $(\gamma, \phi_{c, T_D})=(3.43, 0.8296)$  for $q=7$. In (b): $(\gamma, \phi_{c, R_E})=(2.83, 0.8281)$ for $n=2$, $(\gamma, \phi_{c, R_E})=(2.98, 0.8280)$ for $n=3$, $(\gamma, \phi_{c, R_E})=(3.33, 0.8283)$ for $n=4$.}
\end{figure}

A hallmark of the glass transition is that the dynamics drastically slows down without any evident change in the static pair structure as the liquid approaches its glass transition. We illustrate the emergence of the typical glassy phenomena in mixtures of hard disks and hard ellipses by analyzing the density dependence of structure and dynamics for a system with the disk concentration $x=0.5$ and the ellipse aspect ratio $k=2$. 

In order to characterize the positional and the orientational order for each species, we analyze the pair correlation function and the angular correlation function, defined by
\begin{equation}
	g_{\alpha}(r)=\frac{L^2}{2\pi r\Delta rN_{\alpha}(N_{\alpha}-1)}\left<\sum_{j\neq k}\delta(r-|\textbf{r}_{\alpha,j}-\textbf{r}_{\alpha,k}|)\right>,
\end{equation}
and
\begin{equation}
	g_{2,\alpha}(r)=\left<\cos\{2[\theta_{\alpha}(r)-\theta_{\alpha}(0)]\}\right>,
\end{equation}
where $\textbf{r}_{\alpha,j}$ and $\theta_{\alpha,j}$ designate the position and orientation of particle $j$ belonging to species ${\alpha}$, and the ensemble average $<\cdot\cdot\cdot>$ in Eq. (2) is also performed over all particle pairs with the distance $r$. As shown in Figs. 2(a) and 2(b), the long-range positional order for both species is apparently suppressed even at the highest density studied since $g_{\alpha}(r)$ quickly decays to unity and exhibits several peaks only at short distances. It is also obvious from the splitting of the second peak in $g_{\alpha}(r)$ at high densities that some short-range order develops upon increasing density. This feature is more evident for the disks since their isotropic shape favors the formation of a local hexagonal structure (see Fig. 1). Nevertheless, the overall change in the static pair structure is mild as compared to the dramatic change in the dynamics. The system also lacks any long-range orientational order. The disks are oriented randomly at any density due to their spherical shape and hence, the angular correlation function $g_{2,D}(r)$ for the disks is completely featureless (data not shown). Figure 2(c) indicates that the orientational order of the ellipses increases upon compression. However, even at the highest density investigated, the angular correlation function $g_{2,E}(r)$ for the ellipses decays much faster than that in a nematic liquid crystal,~\cite{Book_deGennes} in which the angular correlation function is expected to decay slower than the power law $g_{2}(r)\sim r^{-1/4}$. Therefore, the mixture remains disordered in both the translational and the rotational degrees of freedom.

In contrast to the weak changes in the static pair structure, the dynamics of glass-forming liquids varies substantially on the approach to the glass transition. The typical behavior of a glass-forming liquid is conventionally described by the dynamic properties such as the relaxation time. The translational dynamics is investigated by the self-intermediate scattering function,
\begin{equation}
	F_{s,\alpha}(q,t)=\frac{1}{N_{\alpha}}\left<\sum_{j=1}^{N_{\alpha}}\exp\{i\textbf{q}\cdot[\textbf{r}_{\alpha,j}(t)-\textbf{r}_{\alpha,j}(0)]\}\right>,
\end{equation}
where $i=\sqrt{-1}$ and $q$ is the wave number, and the rotational dynamics is explored by the $n$th order orientational correlation function,
\begin{equation}
	L_{n,\alpha}(t)=\frac{1}{N_{\alpha}}\left<\sum_{j=1}^{N_{\alpha}}\exp\{in[{\theta}_{\alpha,j}(t)-{\theta}_{\alpha,j}(0)]\}\right>,
\end{equation}
where $n$ is a positive integer. Figure 3 displays $F_{s,\alpha}(q,t)$ at $q=7$ (which is close to the first peak of the static structure at high densities) for both species and $L_{n,E}(t)$ at $n=3$ for the ellipses at various area fractions $\phi$. The translational relaxation dynamics is found to slow down greatly with increasing density for both species. In particular, $F_{s,\alpha}(q,t)$ exhibits a two-step relaxation at sufficiently high area fractions. The two-step decay is characteristic of glassy dynamics and implies two relaxation processes, which are commonly denoted as $\beta$-relaxation and $\alpha$-relaxation.~\cite{Book_Gotze} The $\beta$-relaxation corresponds to the rattling motion of particles being trapped in cages by their nearest neighbors at short times, while the $\alpha$-relaxation reflects the motion of particles escaping from the cages at long times. The rotational dynamics of the ellipses is likewise typical of a glass-forming liquid, as shown in Fig. 3(b). Obviously, similar glassy dynamics does not appear in the rotational dynamics of the disks since their isotropic shape allows them to rotate without hindrance. This implies that the disks always remain ergodic in the rotational degrees of freedom. Our results thus indicate a coupling of the translational dynamics for both species in mixtures of hard disks and hard ellipses, although the structure and the dynamics for different types of particles differ drastically in the rotational degrees of freedom. 

It is instructive to compare the translational dynamics for both species within the same system because such a comparison permits a better understanding for the role of particle shape in the translational motion of particles. Since the particle areas are chosen to be equal for both species, the difference in the translational dynamics is solely induced by the particle shape. Figure 3(a) indicates that the dynamics of the ellipses displays a mild but detectable speed up as compared to that for the disks at low densities [e.g., compare different types of lines at $\phi=0.7$ in Fig. 3(a)]. The same trend holds for high-density states and becomes fairly evident as the liquid enters into the dynamically sluggish regime [see the curves at high densities in Fig. 3(a)]. Thus, the translational motion of particles is promoted by the particle anisotropic shape in the same system, a trend that is more pronounced at higher densities. The same conclusion remains valid for systems with other compositions and other aspect ratios, as discussed in Subsections III B and III C. 

It should be emphasized that the translational dynamics for both species is qualitatively the same at any density despite the presence of quantitative differences. In particular, a single translational glass transition point $\phi_{c,T_{\alpha}}$ is found to occur for both types of particles, where $\phi_{c,T_{\alpha}}$ is determined from the MCT power-law fits of the translational relaxation times $\tau_{T, \alpha}$ for species $\alpha$,
\begin{equation}
	\tau_{T,\alpha}\sim(\phi_{c,T_{\alpha}}-\phi)^{-\gamma}, 
\end{equation}
where $\phi_{c,T_{\alpha}}$ and $\gamma$ are fitting parameters, and $\tau_{T, \alpha}$ is defined as $F_{s,\alpha}(q,t=\tau_{T, \alpha})=0.1$. Figure 4(a) indicates that the $\phi$-dependence of $\tau_{T, \alpha}$ is well described by Eq. (5).~\footnote{The MCT power-law description of the dynamics applies only in a limited range of $\phi$ and the fitted results depend on the $\phi$ range, as found in many other glass formers. Hence, the power-law fits are employed in the present paper to quantitatively characterize the trends of the dynamics in variation with some variable rather than obtain the true glass transition points.} Independent fits produce a common glass transition point $\phi_{c,T_{\alpha}}=0.8290\pm0.0005$ at $q=7$ for both species. Using data for other values of $q$ yields similar glass transition densities, which are given in the caption of Fig. 4. We note, however, that the fitted exponent $\gamma$ depends on $q$ (see the caption of Fig. 4). Thus, the wave number $q$ has merely a quantitative influence on the relaxation times as long as $q$ is chosen to be not far away from the first peak of the static structure, and the fitted glass transition point is independent of $q$. Likewise, we can define the rotational relaxation time $\tau_{R, E}$ for the ellipses as $L_{n,E}(t=\tau_{R, E})=0.1$, and the MCT power law provides a good description for the $\phi$ dependence of $\tau_{R, E}$ at high densities, 
\begin{equation}
	\tau_{R,E}\sim(\phi_{c,R_E}-\phi)^{-\gamma}.
\end{equation}
A common rotational glass transition point $\phi_{c,R_E}=0.8281\pm0.0001$ is obtained for the ellipses using data for various $n$ [Fig. 4(b)]. Again, the fitted exponent $\gamma$ depends on $n$ (see the caption of Fig. 4). Notice that the rotational glass transition sets in at a density quite close to the translational one in Fig. 4, a result that arises because the ellipse aspect ratio is fixed at $k=2$ and that is in agreement with the analysis from the self-diffusion of monodisperse hard ellipses~\cite{JCP_139_024501} and recent measurements for monolayers of colloidal ellipsoids with $k\approx 2$.~\cite{NC_5_3829, PRL_110_188301} The density range between the two glass transitions is found to be nearly independent of mixture's composition but substantially grow with elevating the aspect ratio of the ellipses, as discussed later in detail.

Therefore, we demonstrate that a mixture of hard disks and hard ellipses exhibits typical glass-formation behaviors in the translational degrees of freedom for both types of particles and in the rotational degrees of freedom for the ellipses. The influence of composition and aspect ratio on the glass formation is explored next in Subsections III B and III C.

\subsection{Influence of composition}

\begin{figure}[tb]
	\centering
	\includegraphics[angle=0,width=0.45\textwidth]{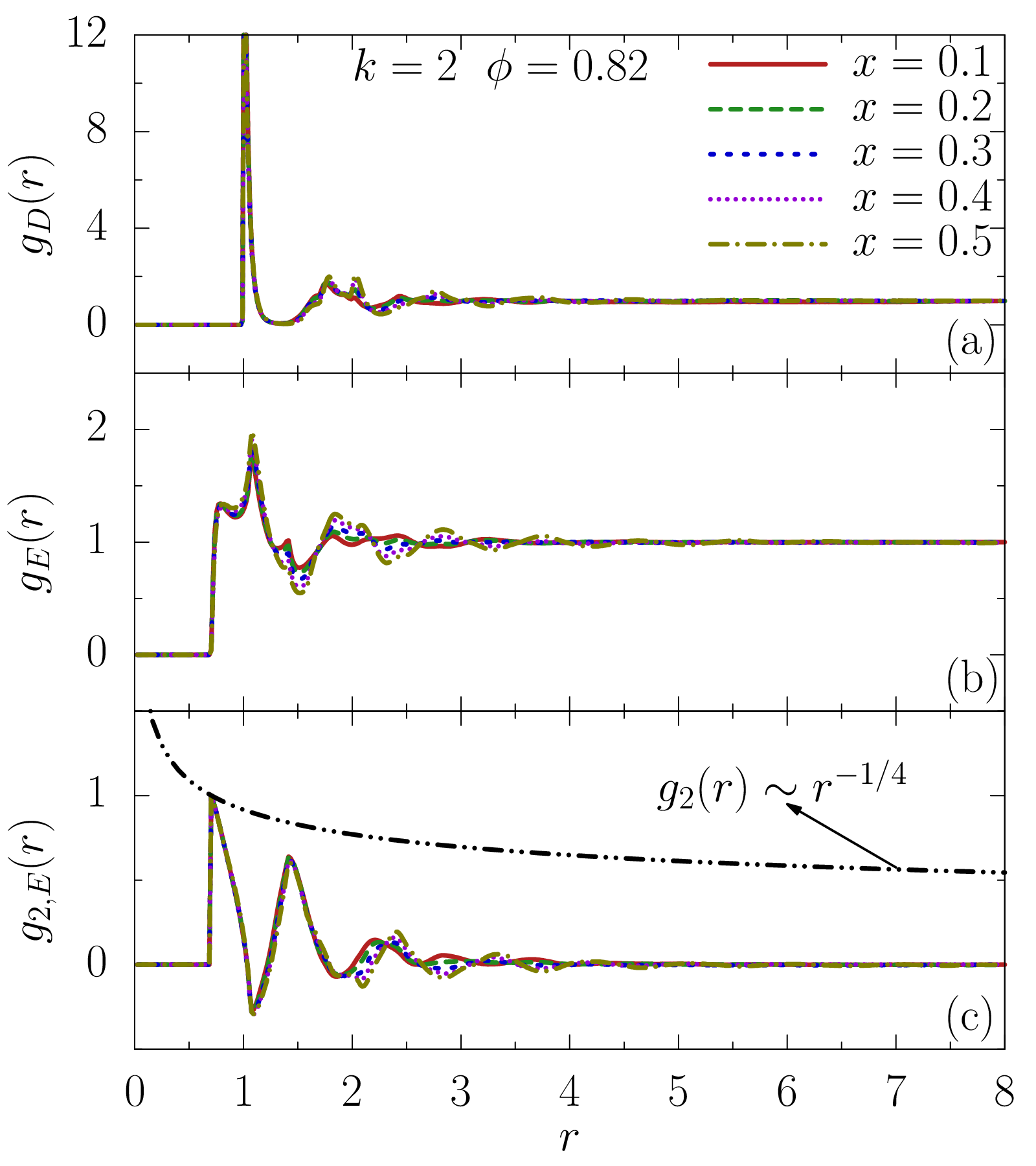}
	\caption{Pair correlation function $g_{\alpha}(r)$ for (a) the disks and (b) the ellipses, and (c) angular correlation function $g_{2,E}(r)$ for the ellipses for mixtures of hard disks and hard ellipses with $k=2$ and various $x$ [indicated in (a)] at $\phi=0.82$. The power law $g_{2}(r)\sim r^{-1/4}$ is included in (c) to indicate the absence of any long-range orientational order in the system.}
\end{figure}

\begin{figure}[tb]
	\centering
	\includegraphics[angle=0,width=0.45\textwidth]{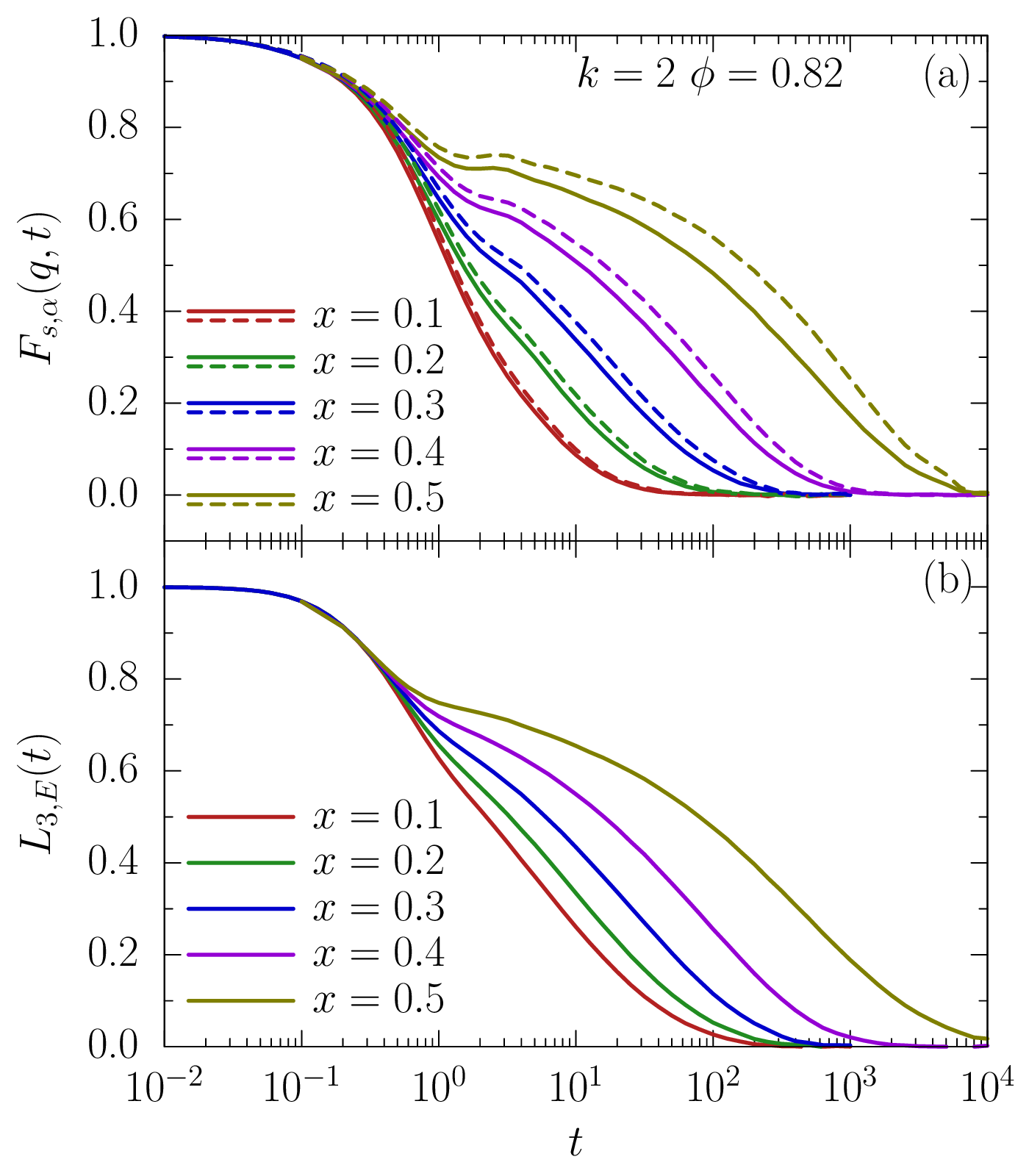}
	\caption{(a) Self-intermediate scattering function $F_{s,\alpha}(q,t)$ at $q=7$ for the disks (dashed lines) and the ellipses (solid lines) and (b) $3$rd order orientational correlation function $L_{3,E}(t)$ for the ellipses for mixtures of hard disks and hard ellipses with $k=2$ and various $x$ at $\phi=0.82$.}
\end{figure}

\begin{figure}[tb]
	\centering
	\includegraphics[angle=0,width=0.45\textwidth]{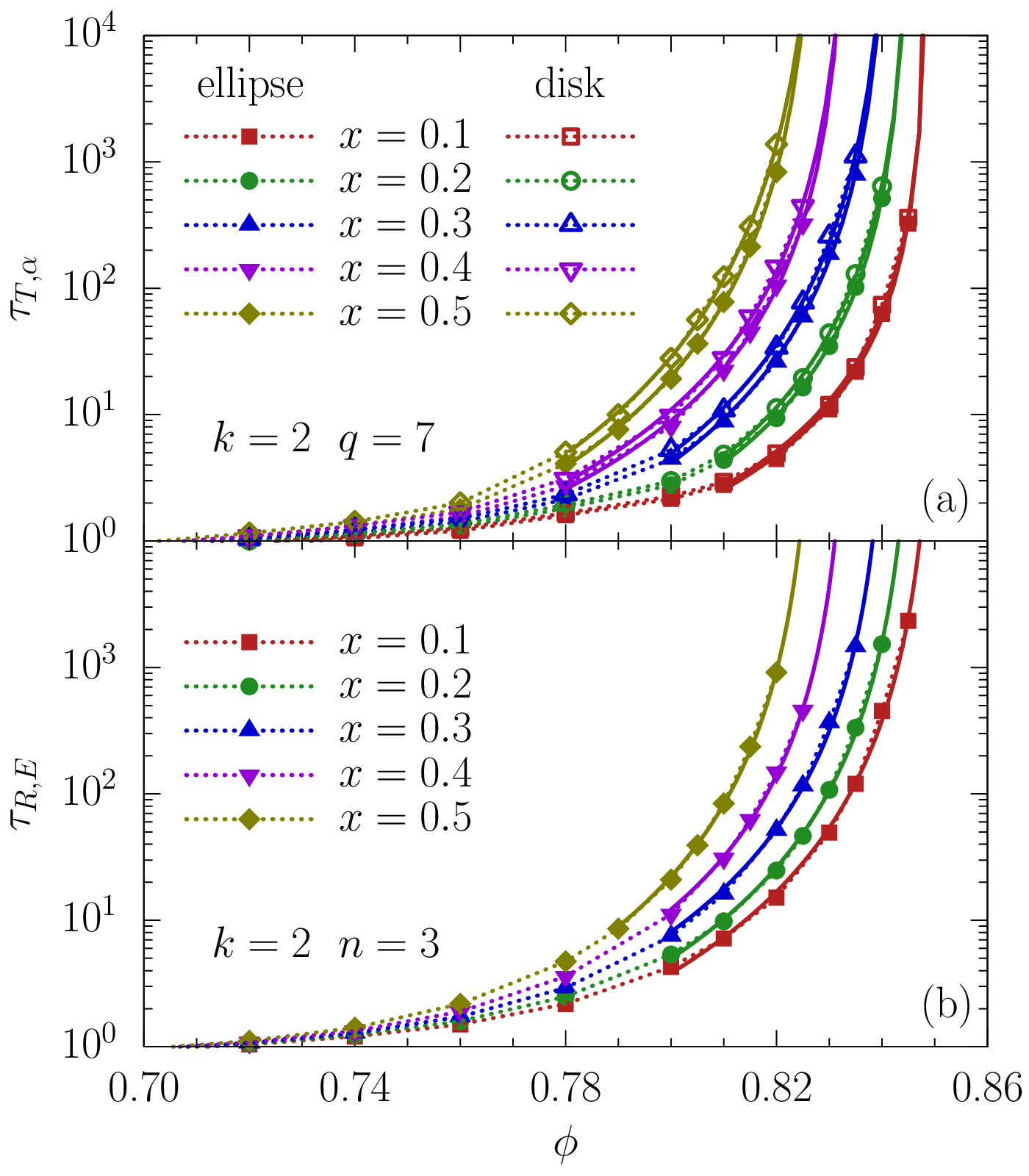}
	\caption{(a) $\tau_{T, \alpha}$ and (b) $\tau_{R, E}$ as a function of $\phi$ for mixtures of hard disks and hard ellipses with $k=2$ and various $x$. Solid lines are MCT fits. The fitted glass transition densities are shown in Fig. 8.}
\end{figure}

\begin{figure}[tb]
	\centering
	\includegraphics[angle=0,width=0.45\textwidth]{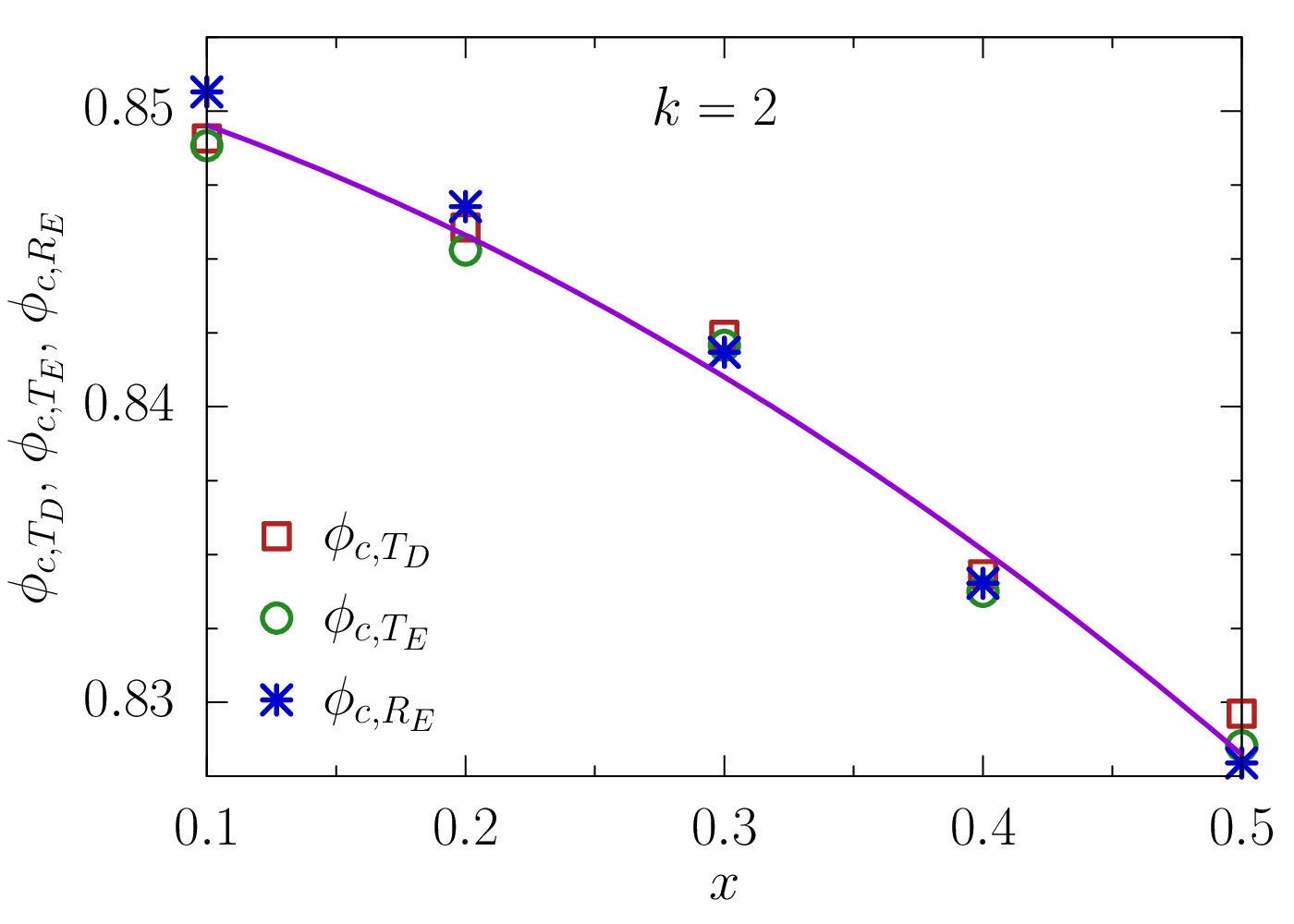}
	\caption{Translational glass transition density $\phi_{c,T_{\alpha}}$ for both species and rotational glass transition density $\phi_{c,R_{E}}$ for the ellipses as a function of $x$ for mixtures of hard disks and hard ellipses with $k=2$. The line is a guide to the eye.}
\end{figure}

Mixture's composition represents a basic parameter for a binary system and has been demonstrated to profoundly affect the glassy dynamics in other types of mixtures, such as binary hard spheres~\cite{PRE_67_021502, PRL_91_085701, PRE_69_011505, JCP_134_054504} and binary hard ellipses.~\cite{SM_11_627} For instance, multiple glasses occur even in simple binary hard-disk mixtures, as indicated by MCT calculations~\cite{EPL_96_36006, PRL_103_205901} and simulations.~\cite{JCP_125_0164507, PRE_74_021409} Here, we demonstrate that the composition strongly influences the glassy dynamics in mixtures of hard disks and hard ellipses. 

We consider several disk concentrations ranging from $x=0.1$ to $0.5$, when the ellipse aspect ratio is fixed at $k=2$. For this aspect ratio, both the positional and the orientational order can be suppressed even when the disk concentration is low (e.g., $x=0.1$), as evidenced by the results in Fig. 5, where we present the pair correlation functions for both species and the angular correlation functions for the ellipses for various compositions $x$ at a fixed area fraction of $\phi=0.82$. Figures 5(a) and 5(b) also reveal that elevating the disk concentration at a fixed density leads to a slight enhancement in the positional order [as reflected by the increasing peak values in $g_{\alpha}(r)$] for both species, in accord with expectations. Dependence of the orientational order for the ellipses on composition is shown to be quite weak [Fig. 5(c)], implying that composition does not play a dominant role in determining the spatial alignment of the ellipses. As shown below, however, altering mixture's composition significantly affects the dynamics of the ellipses in both the translational and the rotational degrees of freedom.

Figure 6 displays the self-intermediate scattering functions for both species and the orientational correlation functions for the ellipses for various compositions $x$ at $\phi=0.82$. Since $q$ and $n$ have merely quantitative effects on the corresponding time correlation functions if they are chosen properly, we focus on $F_{s,\alpha}(q,t)$ at $q=7$ and $L_{n,E}(t)$ at $n=3$ here and in the following. Figure 6(a) clearly indicates that the long-time translational dynamics for both species significantly slows down upon increasing the disk concentration at a fixed density. For example, $F_{s,\alpha}(q,t)$ only exhibits a weak two-step relaxation behavior at $x=0.1$, while the two-step decay becomes fairly evident at $x=0.5$. Accordingly, the translational relaxation time grows by two orders of magnitude when the disk concentration is elevated from $0.1$ to $0.5$ at $\phi=0.82$. Again, the self-intermediate scattering function for the ellipses decays faster than that for the disks in the same system for all compositions studied [compare different types of lines for each $x$ in Fig. 6(a)]. Turning to the rotational dynamics, the rotational motion of the ellipses is found to similarly be influenced greatly by the disk concentration. Specifically, an increase in $x$ similarly leads to a significant slowing down of the long-time rotational dynamics for the ellipses at a fixed density [Fig. 6(b)]. Thus, adding hard disks into a system of hard ellipses tends to inhibit the motion of the ellipses in both the translational and the rotational degrees of freedom.

In order to better illustrate the influence of composition on the dynamics, we present the $\phi$-dependence of the relaxation times for various compositions $x$ in Fig. 7. Both the translational and the rotational relaxation times grow with increasing $x$ at fixed densities, a tendency that becomes more pronounced at higher densities. Figure 7 also illustrates that the relaxation times grow dramatically on the approach to the glass transition and that the high-density data can be well fitted by the MCT power laws (see the solid lines in Fig. 7). Thus, the MCT glass transition densities can be determined from the power-law fits. Again, independent fittings yield a common translational glass transition point for both types of particles for each composition.

Prior studies~\cite{PRE_75_051304, SM_6_2960, PRE_86_041303} indicate that the jamming transition density in a system of pure ellipses with $k=2$ is significantly higher than that in a system of pure disks. Likewise, our recent simulations~\cite{SM_11_627} indicate that the translational glass transition in binary hard ellipses with $k=2$ occurs at a much higher density than that in systems of pure disks. Hence, it is natural to anticipate that elevating the disk concentration leads to a diminished translational glass transition density in the mixture of hard disks and hard ellipses with $k=2$, a trend that is already obvious in Figs. 6 and 7. Such a trend is more clearly illustrated in Fig. 8, where the rotational glass transition density for the ellipses is shown to similarly decrease with increasing the disk concentration. Interestingly, both glass transition densities decrease with the disk concentration at a similar rate and remain close to each other for all compositions studied. We note that the results in Fig. 8 seem to show a slight but consistent dependence of the glass transition densities on composition where the rotational glass transition shifts from higher to lower densities than those corresponding to the translational glass transition. This observation might be due to uncertainties in the fitting procedures and/or arise because the fitted results from the MCT power-law analyses depend on the density range employed. At any rate, however, our analysis indicates that mixture's composition does not significantly affect the density range between the two glass transitions. By contrast, the rotational glass transition for the ellipses sets in at a lower density than the translational one when the ellipses are sufficiently elongated, as discussed in Subsection III C.

\subsection{Influence of aspect ratio}

\begin{figure}[tb]
	\centering
	\includegraphics[angle=0,width=0.45\textwidth]{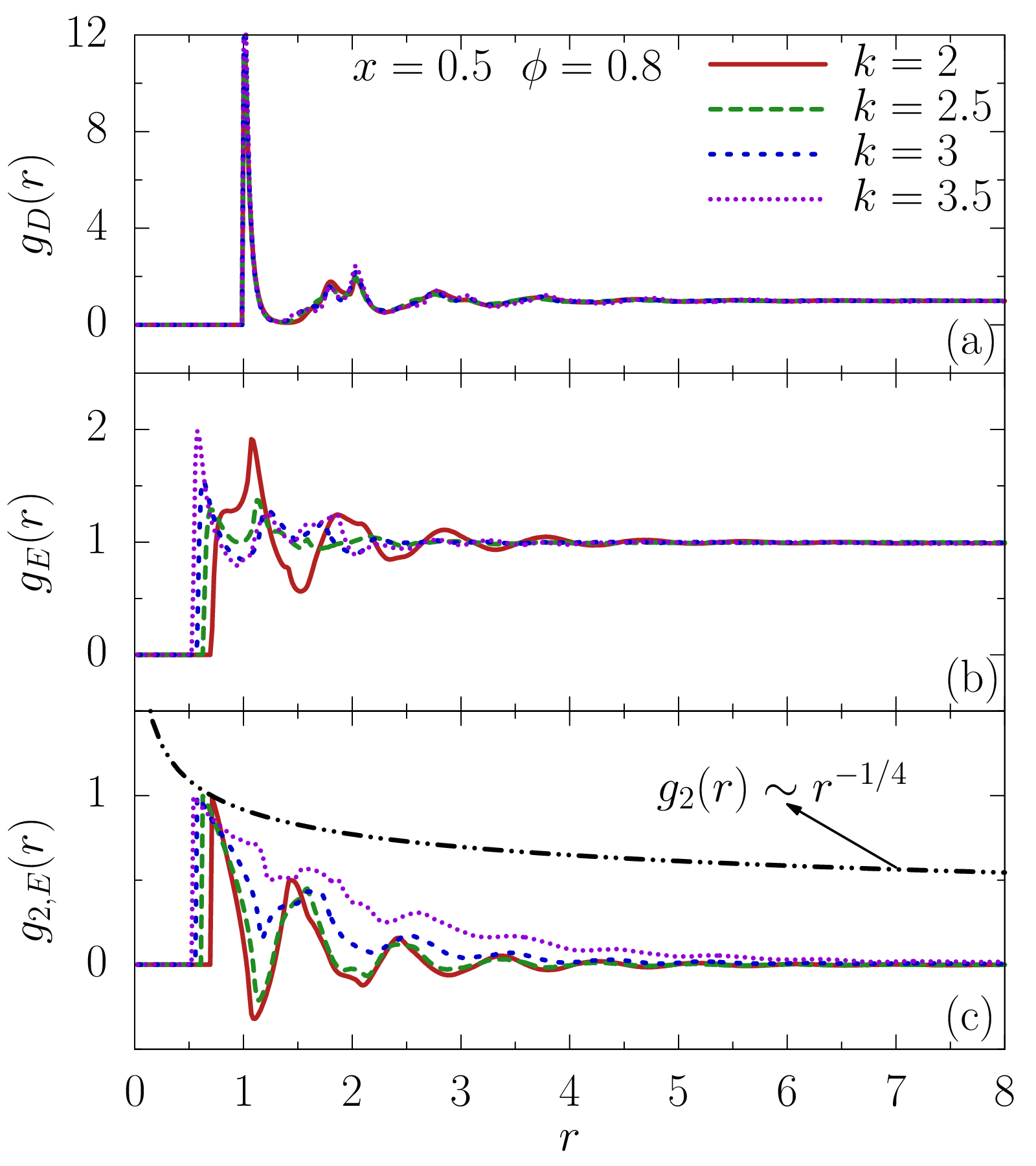}
	\caption{Pair correlation function $g_{\alpha}(r)$ for (a) the disks and (b) the ellipses, and (c) angular correlation function $g_{2,E}(r)$ for the ellipses for mixtures of hard disks and hard ellipses with $x=0.5$ and various $k$ [indicated in (a)] at $\phi=0.8$. The power law $g_{2}(r)\sim r^{-1/4}$ is included in (c) to indicate the absence of any long-range orientational order in the system.}
\end{figure}

\begin{figure}[tb]
	\centering
	\includegraphics[angle=0,width=0.45\textwidth]{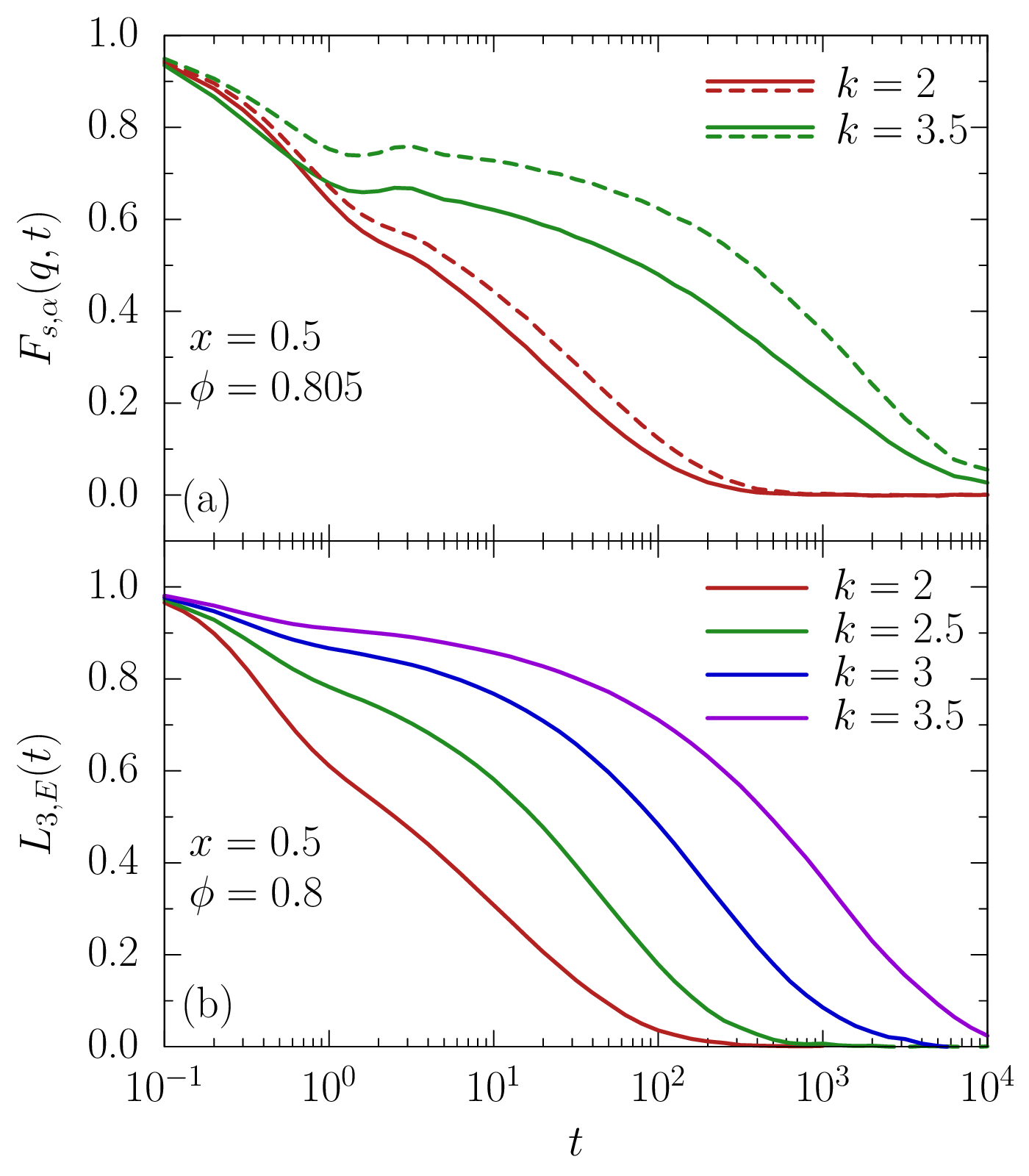}
	\caption{(a) Self-intermediate scattering function $F_{s,\alpha}(q,t)$ at $q=7$ for the disks (dashed lines) and the ellipses (solid lines) and (b) $3$rd order orientational correlation function $L_{3,E}(t)$ for the ellipses for mixtures of hard disks and hard ellipses with $x=0.5$ and various $k$ at a fixed area fraction.}
\end{figure}

\begin{figure}[tb]
	\centering
	\includegraphics[angle=0,width=0.45\textwidth]{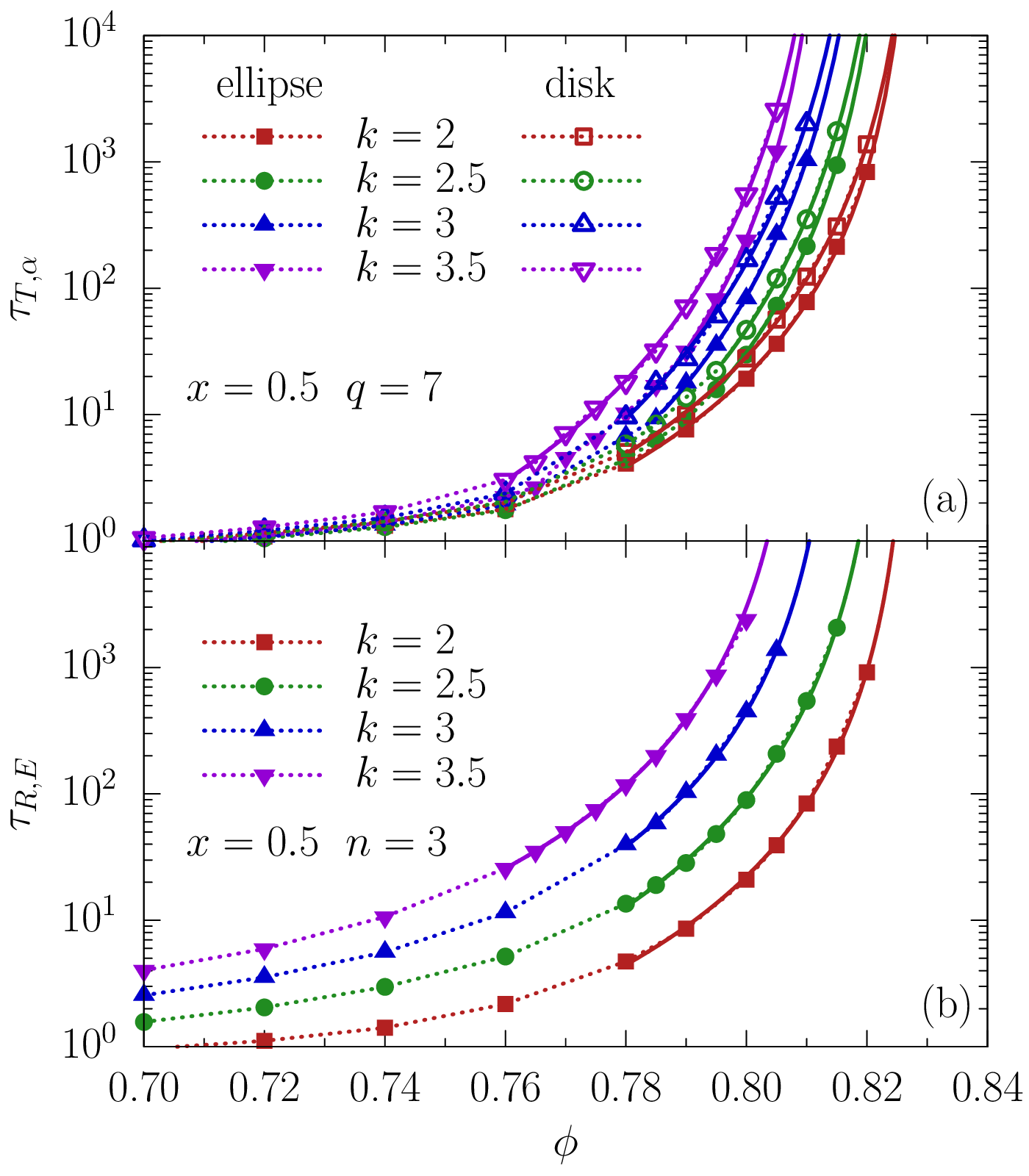}
	\caption{(a) $\tau_{T, \alpha}$ and (b) $\tau_{R, E}$ as a function of $\phi$ for mixtures of hard disks and hard ellipses with $x=0.5$ and various $k$. The solid lines are MCT fits. The fitted glass transition densities are shown in Fig. 12.}
\end{figure}

\begin{figure}[tb]
	\centering
	\includegraphics[angle=0,width=0.45\textwidth]{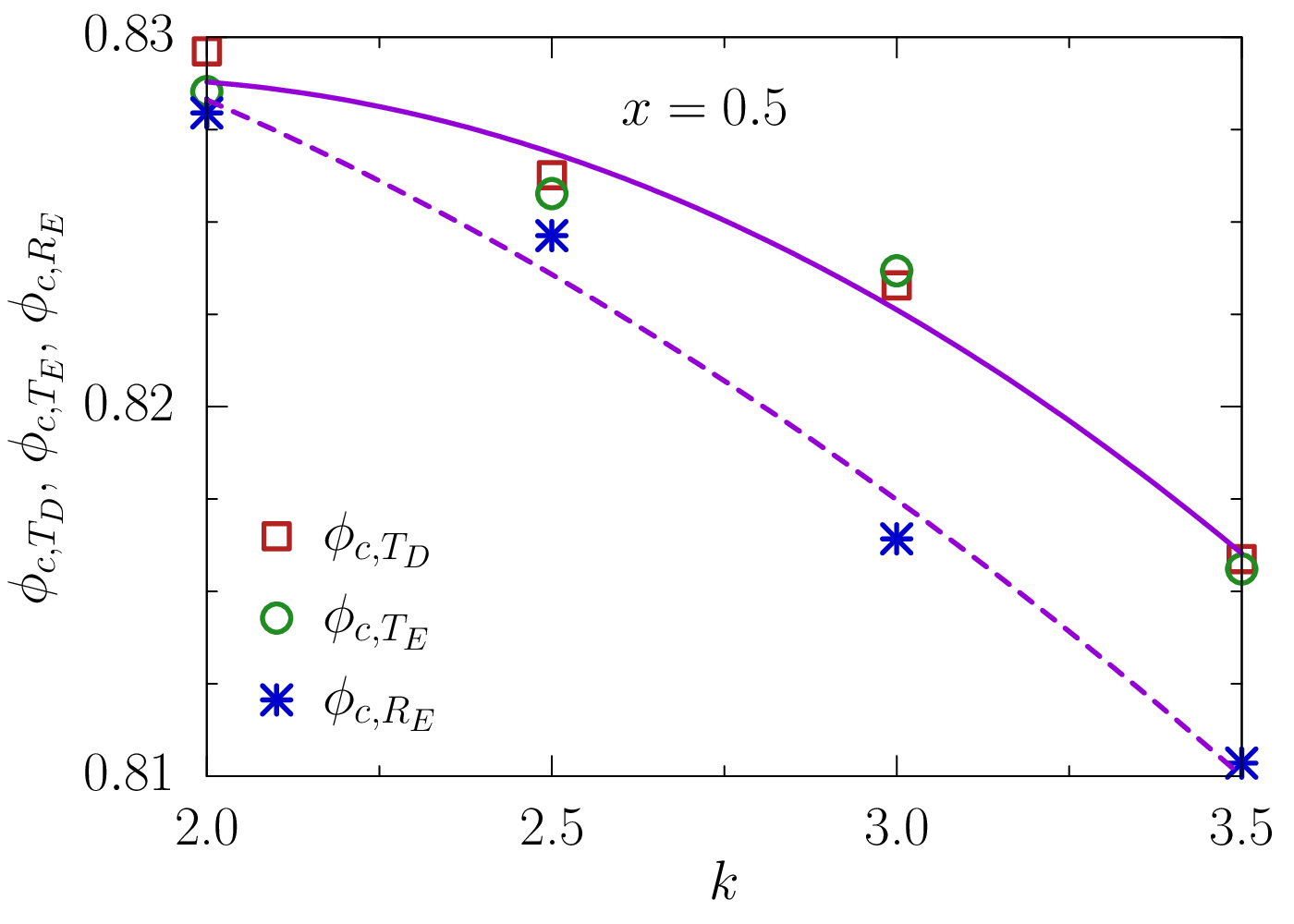}
	\caption{Translational glass transition density $\phi_{c,T_{\alpha}}$ for both species and rotational glass transition density $\phi_{c,R_{E}}$ for the ellipses as a function of $k$ for mixtures of hard disks and hard ellipses with $x=0.5$. The lines are a guide to the eye.}
\end{figure}

We now explore how the ellipse aspect ratio influences the glass formation in mixtures of hard disks and hard ellipses. For comparative purposes, we fix the disk concentration at $x=0.5$. In principle, the influence of aspect ratio becomes more evident as the ellipses become more elongated and hence, it is crucial to investigate the glassy dynamics for mixtures where the ellipse aspect ratio is large enough. However, we have found that different types of particles exhibit a strong tendency to separate from each other at high densities for large aspect ratios (e.g., $k\geq 4$), a behavior that complicates the analysis of glass formation at any rate. On the other hand, if the aspect ratio is too small (e.g., $k\leq1.5$), the rotational dynamics of the ellipses resembles that of a free rotator, which is quite different from the typical glassy dynamics and which has been also observed in hard ellipsoids with very small aspect ratios.~\cite{PRL_98_265702, EPL_84_16003} The above considerations restrict our attention to the mixtures with $2\le k\le 3.5$, where disks and ellipses are well mixed even at the highest densities investigated and the ellipses exhibit typical glassy dynamics upon increasing density.

We first discuss the influence of aspect ratio on the static structure of the system. Figure 9(a) indicates that the static pair structure for the disks is only weakly influenced by the aspect ratio of the ellipses. The positional order of the ellipses displays a slightly diminished trend as the aspect ratio elevates [Fig. 9(b)], in line with the previous finding~\cite{JCP_139_024501} that shape anisotropy tends to weaken the spatial correlations of particles in the translational degrees of freedom. In contrast, alterations in the aspect ratio \textcolor[rgb]{1,0,0}{lead} to evident changes in the structure of the ellipses; increasing the aspect ratio enhances the orientational order of the ellipses [see Fig. 9(c)], because the particle alignment is entropically more favored for systems composed of more elongated particles. Notice that the ellipses do not form a nematic phase even for the largest $k$ studied, since the angular correlation function still decays much faster than that for a nematic liquid crystal. Thus, the ellipses undergo a rotational glass transition at sufficiently high densities.

Turning to the dynamics, Fig. 10 exhibits the time correlation functions for various aspect ratios $k$ at a fixed density. Since the rotational motion of the ellipses is expected to become more hindered for larger aspect ratios, increasing $k$ inevitably leads to a slowing down of the rotational dynamics for the ellipses at fixed densities [Fig. 10(b)]. In fact, the influence of the aspect ratio on the rotational dynamics of the ellipses is significant even at low densities [see Fig. 11(b)], in accord with the previous analysis on the basis of diffusive properties for hard ellipses.~\cite{JCP_139_024501} When the aspect ratio $k$ is elevated from $2$ to $3.5$, the translational dynamics also slows down for both species at a fixed density and the difference in the dynamics between the two species grows within the same system. Since the high-density dynamics traces the glass transition point, the trend in Fig. 10(a) appears to be consistent with the previous studies on the jamming transition in two-dimensional ellipsoidal particle systems,~\cite{PRE_75_051304, SM_6_2960, PRE_86_041303} demonstrating that the jamming transition density first grows quickly with increasing $k$, achieves a maximum at a moderate aspect ratio ($k\approx 1.5$), and then slowly becomes smaller for larger aspect ratios.

Figure 11 depicts the $\phi$-dependence of the relaxation times for various aspect ratios $k$. The influence of the ellipse aspect ratio on the translational dynamics is more pronounced at higher densities, while the particle shape plays a dominant role in determining the rotational dynamics of the ellipses even at low densities. Both the translational and the rotational relaxation times grow drastically on approaching the glass transition. Again, we determine the glass transition points from the MCT power-law fits in order to make a quantitative analysis for both glass transitions. The variations of the fitted glass transition densities with the ellipse aspect ratio are displayed in Fig. 12. In line with the above analysis for the dynamics in Fig. 10, we observe that both the translational and rotational glass transition densities diminish as the ellipse aspect ratio elevates. Figure 12 further reveals another result: The rotational glass transition density decreases with the aspect ratio at a faster rate than the translational one. Consequently, a density regime, where an orientational glass for the ellipses forms, emerges due to elevations in the ellipse aspect ratio. This regime extends with $k$, a trend that is in good agreement with recent experiments~\cite{PRL_107_065702, NC_5_3829} for monolayers of colloidal ellipsoids and MCT predictions.~\cite{PRE_62_5173} Therefore, the ellipse aspect ratio determines the density range where an orientational glass forms for the ellipses in mixtures of hard disks and hard ellipses.

\section{Summary}

In summary, we explore the glassy dynamics in binary mixtures composed of hard disks and hard ellipses. The area of the ellipses is chosen to be identical to that of the disks, a choice that facilitates elucidating the role of particle shape in determining the dynamics since the dynamics for both species can be compared within the same system at the same particle size. Glass formation is demonstrated to occur in the translational degrees of freedom for both types of particles and in the rotational degrees of freedom for the ellipses under appropriate conditions. The translational dynamics for the ellipses is found to be faster than that for the disks within the same system. Both types of particles display a common translational MCT glass transition point despite the presence of quantitative differences in their dynamics. We assess the influence of composition and aspect ratio on the glass formation. Our results indicate that both glass transition densities are strongly affected by mixture's composition. Nevertheless, the density range between the two glass transitions appears to be independent of composition, because both glass transition densities decrease at a similar rate as the disk concentration increases. The ellipse aspect ratio is shown to profoundly impact on the glass formation in mixtures of hard disks and hard ellipses. The rotational glass transition density diminishes at a faster rate than the translational one when the aspect ratio is elevated. Therefore, the density range, where an orientational glass forms for the ellipses, appears for sufficiently large aspect ratios. Our simulations thus imply that mixtures of particles with different shapes emerge as a promising model for exploring the role of particle shape in determining the properties of glass-forming liquids. Our work also illustrates the potential of using knowledge concerning the dependence of glass-formation properties on mixture's composition and particle shape to assist in the rational design of amorphous materials.

\begin{acknowledgments}
This work is supported by the National Basic Research Program of China (973 Program, 2012CB821500), and the National Natural Science Foundation of China (21474111, 21222407, 21404103).
\end{acknowledgments}

%\appendix

%\bibliographystyle{apsrev4-1}
\bibliography{refs}% Produces the bibliography via BibTeX.

\end{document}